\documentclass[aps,prb,twocolumn,showpacs,floatfix]{revtex4}

\pdfoutput=1

\usepackage{ulem}
\usepackage{times}
\usepackage{amsfonts}
\usepackage{graphicx}
\usepackage{pst-all}
\usepackage{graphics}
\usepackage{amssymb}
\usepackage{layout}
\usepackage{verbatim}
\usepackage{amsfonts}
\usepackage{epsfig}
\usepackage{bm}
\usepackage{amsmath}
\usepackage{graphicx}
\usepackage{hyperref}
\usepackage{epsf}
\usepackage{bm}
\usepackage{verbatim}

\newcommand \bea {\begin{eqnarray} }
\newcommand \eea {\end{eqnarray}}

\newcommand{\beg}{\begin{equation}}
\newcommand{\en}{\end{equation}}

\newcommand \bel  {\begin{align}}
\newcommand \enl  {\end{align}}

\newcommand{\dg}{^{\dagger }}
\newcommand{\bk}{\mathbf k}
\newcommand{\hbk}{\hat{\mathbf k}}

\newcommand{\bfr}{\mathbf r}

\newcommand{\br}{\mathbf r}
\newcommand{\bR}{{\bf{R}}}

\newcommand{\up}{\uparrow}
\newcommand{\dn}{\downarrow}

\begin{document}
\title{{A Theory of Topological Kondo Insulators}}
\author{Maxim Dzero$^{1}$, Kai Sun$^2$, Piers Coleman$^{3,4}$ and Victor Galitski$^2$}
\affiliation{$^{1}$ Department of Physics, Kent State University, Kent, OH 44242, USA\\
$^2$ Joint Quantum Institute and Condensed Matter Theory Center, Department of Physics, University of Maryland, College Park, MD 20742, USA\\
$^3$Center for Materials Theory, Rutgers University, Piscataway, NJ 08854, USA\\
$^4$Department of Physics, Royal Holloway, University of London, Egham, Surrey 
TW20 0EX, UK }

\date{\today}

\begin{abstract}
We examine how the properties of the Kondo insulators change when the
symmetry of the underlying crystal field multiplets is taken into
account.  We employ the Anderson lattice model and consider its
low-energy physics.  We show that in a large class of crystal field
configurations, Kondo insulators can develop a topological non-trivial
ground-state. Such topological Kondo insulators are adiabatically
connected to non-interacting insulators with unphysically large
spin-orbit coupling, and as such may be regarded as interaction-driven
topological insulators.  We analyze the entanglement entropy of the
Anderson lattice model of Kondo insulators by evaluating its
entanglement spectrum. Our results for the entanglement spectrum are
consistent with the surface state calculations.  Lastly, we discuss
the construction of the maximally localized Wannier wave functions for
generic Kondo insulators.
\end{abstract}


\maketitle \section{Introduction} Topological insulators are a novel
class of materials in which strong spin-orbit interaction leads to the
inversion of the band gap (See Ref.~\onlinecite{Hasan2010} and
\onlinecite{{Qi2010}} and references therein). In 3D, this inversion
results in chiral metallic surface states due to a formation of a
single Dirac cone inside the gap \cite{Fu2007, Moore2007, Roy2009,
Hsieh2008, Xia2009}.  Among materials which exhibit this behavior,
for example, HgTe, Bi$_2$Se$_3$, Bi$_{1-x}$Sb$_x$, Bi$_2$Te$_3$ and
TlBiTe$_2$, the chiral structure of the surface states has been
confirmed experimentally~\cite{exp1,exp2,exp3,exp4}.

The emergence of surface modes in topological insulators 
is a band structure effect which can be understood without
invoking interactions.  There is great current interest in the
possibility of interaction driven topological phenomena.
Up until now there are no experimental examples of interaction-driven
topological insulators that preserve time-reversal symmetry. 
However, several theoretical proposals have
been put forward: 2D topological insulators via spontaneous symmetry
breaking in bilayer graphene and optical lattice
systems~\cite{Raghu2008, Sun2009, Nandkishore2010, Sun2010},
topological Mott insulating phase in Ir-based pyrochlore oxides
A$_2$Ir$_2$O$_7$ with A=Nd,Pr \cite{pyro1,pyro2,pyro3,pyro4}, Kondo
insulators with the most salient example of SmB$_6$ \cite{Dzero2010}
and insulating behavior in filled skutteridites \cite{skuter}. In this paper
we focus on general principles governing the emergence of
chiral metallic states in Kondo insulators. Throughout this paper, we
use the term ``Kondo insulator'' in its broadest sense, 
including both mixed valent materials 
\cite{Chazalviel1976} such as SmB$_6$ and YbB$_{12}$ and those  in the
more localized limit, such as Ce$_3$Bi$_4$Pt$_3$.

Kondo insulators are a type of heavy fermion
material, first discovered forty years ago \cite{Geballe69}, in which
highly renormalized $f$-electrons hybridize with conduction
electrons to form a completely filled band of quasiparticles with
excitation gaps in the millivolt range
\cite{KIReviews1,KIReviews2,KIReviews3,Coleman2007}.  Because Kondo
insulators appear as a result of strong interactions, one might think
that their excitations and their ground-states are adiabatically
connected to trivial non-interacting band insulators \cite{allen}.  {However,
before jumping to this conclusion one needs to be careful, for in the
renormalization process the width of the heavy electron bands drops 
far below the characteristic size of the spin-orbit interactions, 
driving the physics to a new fixed point characterized by infinite
spin-orbit coupling in the localized bands. Indeed, we shall show that
topological Kondo insulators are adiabatically connected to non-interacting
topological insulators with an unphysically large value of the 
spin-orbit coupling, and in this sense, they are interaction-drive insulators.}

One of the most important features of the $f$-electron systems in
general and Kondo insulators in particular is that the $f$-electron
states are classified with respect to their momentum $\bk$, total
angular momentum $J$ and its $z$-axis component $M$, while conduction
electron states are described by a momentum and a spin $\sigma$.
When an f-electron escapes into the conduction sea, it hybridizes with
a spin-orbit coupled Wannier state of the conduction electrons that
has the same symmetry as the f-state. 
The spin-orbit coupled Wannier
states of the conduction electrons are then decomposed in terms of
plane-wave states and this gives rise to momentum-dependent form
factors with symmetries 
that are uniquely determined by the local symmetry of the f-states.
In this way, the form-factors 
encode the effect of the strong spin-orbit coupling. More
importantly, these form factors also define the underlying symmetry of the
hybridization amplitude and gap which is develops below the ``Kondo
temperature'' $T_K$ at which heavy quasi-particles develop. 
One of the key properties of the spin-orbit coupled f-state,
is an odd-parity wavefunction.  It is the protected odd-parity of the f-states
that provides the driving force for the formation of topological
insulating states. 

The dimension of the form factor matrix is determined by the degeneracy of
the underlying ground state $f$-ion multiplet. In the crystalline
environment the $(2J+1)$ multiplet degeneracy is lifted by the
crystalline fields.  For half-integer values of $J$ the lowest
possible degenerate multiplet is a Kramers doublet, which means that
the form factor is a two-dimensional matrix. For the
integer values of $J$ the crystal field can fully lift degeneracy of
the multiplet. 
This situation corresponds to the non-magnetic state of the $f$-ion 
and currently there are no known examples of such Kondo insulators. 
Thus, in this article we will only focus on the magnetic ions 
with half-integer values of the total angular momentum.

As mentioned above, the symmetry of the lowest lying multiplet
determines the symmetry of the hybridization amplitude. Generically,
two possible scenarios can arise depending on whether the hybridization
contains nodes or not.
For a small, but important subset of these systems, the hybridization 
contains nodes. In this case, the Kondo insulating state is replaced by a 
heavy
semi-metal with a pseudogap, as in the case of CeNiSn or CeRhSb. 
If the nodes correspond to touching of the two non-degenerate bands with 
linear dispersion, 
the system becomes a Weyl semi-metal, 
where topologically-protected surface modes emerge~\cite{wang2011}. 
Note, that the lifting of degeneracy can only happen due to onset of magnetic order. 
The magnetic moments may appear as a result of incomplete screening similarly to what 
happens in CeCoIn5, for example. 

In our previous work, we demonstrated within a mean-field model, that in the 
large class of systems without nodes in the hybridization gap, 
Kondo insulators can develop topological
insulating ground-states\cite{Dzero2010}.  In this paper we develop
this idea in detail, providing mathematical details of the
construction of the wavefunction and explicitly computing the
surface modes for a topological Kondo insulator. 

Although attempts to establish the general principle determining the
relative position of the crystal field multiplets have been made
\cite{Kikoin1994,Ikeda1996,Moreno2000}, but no such principle has yet
been discovered. Experimentally, however, the symmetry of the lowest
lying as well as excited multiplets can be detected, for example, by
inelastic neutron scattering spectroscopy \cite{Rob}. Nevertheless, by
assuming a specific symmetry of the ground state multiplet one is able
to theoretically predict the physical properties of a Kondo
insulator. In addition, it provides multiple ways to verify them
experimentally. To be more precise, the presence of the chiral states
on the surface of Kondo insulators will allow one to indicates
unambiguously the symmetry of the lowest lying multiplet.

Apart from going beyond the brief description of topological Kondo
insulators reported in Ref. \onlinecite{Dzero2010}, we will
discuss the topological properties of the eigenfunctions of the model
Hamiltonian describing the Kondo insulators.  We will start with a
short review of the model and recently obtained results.  We evaluate
the entanglement spectrum for simplest model of the Kondo insulator
corresponding to the nearest neighbors tight-binding approximation for
the conduction bands and a Kramers doublet. We also discuss the choice
of the proper basis for construction of the maximally localized
Wannier functions for the Kondo insulators. We show that Wannier
functions can be constructed on the basis composed of the linear
combination between the conduction and $f$-electron states. Finally,
we provide a short review of available experimental data which points
towards the existence of the chiral surface states in Kondo
insulators.

\section{Anderson Lattice Model}
We begin with writing down the model Hamiltonian to describe the physics of the Kondo insulators. 
In what follows we will consider the
most general case by assuming that there are $N_c$ conduction bands, so that the Hamiltonian 
describing conduction electrons is 
\beg\label{Hc}
H_c=\sum\limits_{l=1}^{N_c}\sum_{\bk , \sigma }\xi_{l\bk }c^{(l)\dagger}_{\bk \sigma }c_{\bk \sigma }^{(l)}
\en
where $\xi_{l\bk}$ is the dispersion of the $l$th band of
conduction electrons, $\sigma$ is the spin index 
and $c_{\bk\sigma}^{(l)\dagger}$ is a conduction electron creation operator. 
Consequently, the Hamiltonian which describes the $f$-electrons is:
\beg\label{Hf}
H_f=\sum\limits_{j}\sum\limits_{\alpha=1}^{N_{\Gamma}}\epsilon_{f\Gamma} f_{j\alpha}\dg f_{j\alpha}+{U}\sum\limits_{i\alpha\alpha'}
f_{i\alpha}\dg f_{i\alpha}f_{i\alpha'}\dg f_{i\alpha'}.
\en
where $f_{j\alpha}\dg$ creates an $f$-electron on site $j$ in a state 
$\alpha$ of a lowest lying multiplet $N_\Gamma$-degenerate multiplet denoted by $\Gamma$ (see below), 
$\epsilon_{f}$ is the $f$-electron energy and $U>0$ is the strength of the Hubbard interaction between the $f$-electrons. We emphasize that index $\alpha$ is not a spin index due to the presence of the strong spin-orbit coupling.  Generally states belonging to the
multiplet $\Gamma$ are described by the total angular momentum $J$ and $z$-component $M$ or some linear superposition of those states and in the second term of Eq. \eqref{Hf}, the summation is restricted to $\alpha\not=\alpha'$.

Finally the term describing how electrons in $N_c$ conduction bands are hybridized with localized $f$-electrons is
\beg\label{Hh}
H_h=\sum\limits_{l=1}^{N_c}\sum
\limits_{j,\alpha=1}^{N_\Gamma}\left[V_{i\sigma,j\alpha}^{(l)} {c}_{i\sigma}^{(l)\dagger} {f}_{j\alpha}+ {\rm h.c.}\right],
\en
Here $V_{i\sigma,j\alpha}^{(l)}$ is a non-local hybridization matrix element 
between the conduction electrons in $l$th band and localized $f$-electrons. 
Thus, the periodic Anderson model Hamiltonian, which is the basis for our subsequent discussion, reads:
\beg\label{HPALM}
H_{PAM}=H_c+H_f+H_h
\en

The hybridization matrix elements $V_{i\sigma,j\alpha}^{(l)}$ can be written as follows :
\beg\label{Vij}
V _{ i\sigma,j\alpha}^{(l)}=V_l\sum_{\bk\sigma }[\Phi_{\Gamma\bk}]_{\alpha \sigma }
e^{i \bk \cdot({\bf R}_i-{\bf R}_j)},
\en
where $V_l$ is the hybridization amplitude and the form factors $[\Phi_{\Gamma\bk }]_{\alpha \sigma }$ are $(2J+1)\times 2$ dimensional matrices given by:
\beg\label{Phias}
[\Phi_{\Gamma\bk }]_{\alpha \sigma }=\langle k\Gamma \alpha \vert \bk
\sigma \rangle
\en
Since in this paper we will be discussing the materials when an $f$-ion is in the valence state with $J=5/2$ 
($f^1$ for cerium or $f^3$ for samarium) it follows
\begin{equation}\label{eq2}
[\Phi_{\Gamma\bk }]_{\alpha \sigma }=\sum_{m\in [-3,3]}\left\langle \Gamma\alpha \Bigl\vert 3m,
\frac{1}{2}\sigma
\right\rangle \tilde{Y}^{3}_{m-\sigma } ({\bk} )
\end{equation}
and 
\begin{equation}\label{}
\tilde{Y}^{3}_{M} (\bk )= \frac{1}{Z}\sum_{\bf R \neq 0} Y^{3}_{M} (\hat {\bf
R}) e^{i \bk\cdot {\bf R}}
\end{equation}
is a tight-binding generalization of the spherical Harmonics that
preserves the translational symmetry of the hybridization,
$\Phi (\bk )= \Phi (\bk +{\bf G})$,  where $\bf{G}$ is 
reciprocal lattice vector.  Here, 
 $\bf R$ are the positions of the Z nearest neighbor sites
around the magnetic ion. Note, that deriving (\ref{Phias}) we have assumed that the symmetry
of the conduction electron amplitude coincides with the symmetry of the $f$-ion multiplet 
\cite{Coqblin1969,Flint2008}. Consequently, we treat the system with only one hybridization channel. 
Now let us recall the definition of the form factors: 
\beg\label{FF}
[\Phi_{\hbk}]_{\alpha\sigma}=\sum\limits_{M=-5/2}^{5/2}\langle k\alpha|JM\rangle\langle JM|\hbk\sigma\rangle,
\en
where $\langle JM|\hbk\sigma\rangle$ is a $(2J+1)\times 2$ matrix whose elements are given by 
$\sigma\sqrt{\frac{\frac{7}{2}-M \sigma }{7}}Y^{3}_{M-\frac{1}{2}\sigma } (\hbk )$. The elements of the matrix
$\langle k\alpha|JM\rangle$ are determined by the specific choice of the $f$-ion multiplet and the corresponding
wave-functions denoted by $|\Gamma\alpha\rangle$. As we have already mentioned above, we will focus our discussion on the case
of $f$-ion with $J=5/2$. This situation is relevant for all known $f$-electron Kondo insulators. Consequently,  in a cubic crystal field environment, the magnetic ion multiplet is split into a doublet 
\beg\label{doub}
\begin{split}
&|\Gamma_1^{(c)}\pm\rangle=\pm\sqrt{\frac{5}{6}}\left\vert\pm\frac{3}{2}\right\rangle\mp\sqrt{\frac{1}{6}}\left\vert\mp\frac{5}{2}\right\rangle
\end{split}
\en
and a quartet
\beg\label{quart}
\begin{split}
|\Gamma_2^{(c)}\pm\rangle&=\pm\sqrt{\frac{1}{6}}\left\vert\mp\frac{3}{2}\right\rangle\pm\sqrt{\frac{5}{6}}\left\vert\pm\frac{5}{2}\right\rangle, \\
|\Gamma_3^{(c)}\pm\rangle&=\pm\left\vert\pm\frac{1}{2}\right\rangle,
\end{split}
\en
so that for this case the matrix $\langle k\alpha|JM\rangle$ is
\begin{eqnarray}\label{overlap1}
&&\langle k \alpha_i \vert J M \rangle= \\
 &=& \ \ \ \ \hskip 0.03in \left(\begin{matrix}
0 & -\sqrt{\frac{1}{6}} & 0 & 0 & 0 & -\sqrt{\frac{5}{6}}\cr
0 & 0 & 1 & 0 & 0 & 0 \cr
0 & 0 & 0 & -1 & 0 & 0\cr
\sqrt{\frac{5}{6}} & 0 & 0 & 0 & \sqrt{\frac{1}{6}} & 0 \cr 
0 & \sqrt{\frac{5}{6}} & 0 & 0 & 0 & -\sqrt{\frac{1}{6}} \cr
\sqrt{\frac{1}{6}} & 0 & 0 & 0 & -\sqrt{\frac{5}{6}} & 0 \cr
\end{matrix}
 \right)
 \end{eqnarray}

In a tetragonal crystal field environment, relevant for
Ce-based Kondo insulators, the Ce multiplet is split into three doublets: 
\begin{equation}
\begin{split}
&|\Gamma_1^{(t)}\pm\rangle=|\pm1/2\rangle, \\
&|\Gamma_2^{(t)}\pm\rangle=\cos(\beta)|\mp3/2\rangle+\sin(\beta)|\pm
5/2\rangle, \\
&|\Gamma_3^{(t)}\pm\rangle=\sin(\beta)|\mp3/2\rangle-\cos(\beta)|\pm
5/2\rangle,
\end{split}
\end{equation}
 where the mixing angle $\beta$ defines orientation of the
corresponding states. 
In an orthorhombic environment, the Kramer's
doublets are generally described by a linear superposition of all three
wave-functions \cite{Kikoin1994,Moreno2000} 
\begin{equation}
|\Gamma^{(ortho)}\pm\rangle= u\vert
\pm 1/2\rangle+v\vert\mp 3/2\rangle+w\vert \pm 5/2 \rangle.
\end{equation} 
Having provided the scheme for the computation of the form-factors we proceed with the discussion of the low-energy
properties of our model (\ref{HPALM}) of Kondo insulators. 

\section{low-energy theory for C\lowercase{e}-based Kondo insulators}
The low-energy properties of the model (\ref{HPALM}) are described in terms of
renormalized quasiparticles formed via strong hybridization between the $c-$ and
$f-$ states and on-site repulsion $U$ between the $f$-electrons. 
In the regime where the $f$ states are predominantly localized, $U\sim W$ ($W$ is the bandwidth), 
we can neglect the momentum dependence of the $f$-electron self-energy
$\Sigma_f(\bk,\omega)\simeq\Sigma_f(\omega)$. 

Below we discuss the topological properties of the effective low-energy model. To make our discussion more
tractable, we will consider separately several experimentally relevant cases. In what follows we 
discuss the simplest case of the single conduction band and Kramers doublet as a ground state multiplet of the magnetic ion. This is done with an eye toward the transport experiments on the Ce-based Kondo insulators \cite{Hundley1990,CeNiSn2002}.

\subsection{single conduction band hybridized with the Kramers doublet: Ce-based Kondo insulators}
In order to derive an effective low-energy model for Kondo insulators, we first introduce the following 
correlation functions for $c$- and $f$-electrons:
\beg
\begin{split}
G_{cc}(\bk,\tau)&=-\langle\hat{T}_\tau\{c_{\bk\sigma}(\tau)c_{\bk\sigma}\dg(0)\}\rangle, \\
G_{ff}(\bk,\tau)&=-\langle\hat{T}_\tau\{f_{\bk\alpha}(\tau)f_{\bk\alpha}\dg(0)\}\rangle, \\
\end{split}
\en
By writing down equations of motion for the $c$-operators with the Hamiltonian (\ref{HPALM}) and going into Matsubara frequency representation we derive the following relation:
\beg
G_{cc}(\bk,i\omega)=G_{cc}^{(0)}(\bk,i\omega)+\frac{|V|^2\Delta_\bk^2}{(i\omega-\xi_\bk)^2}G_{ff}(\bk,i\omega)
\en
with $\Delta_\bk^2=\frac{1}{2}\text{Tr}[{\Phi}_{\Gamma\bk}\dg{\Phi}_{\Gamma\bk}]$ and $G_{cc}^{(0)}(\bk,i\omega)$ is a conduction
electron propagator in the absence of interactions. If we denote the $f$-electron
self-energy by $\Sigma_f(\bk,\omega)$ and keep in mind that this self-energy appears as a result of Hubbard correlations only, 
then it follows:
\beg\label{Gff}
G_{ff}(\bk,i\omega)=\left[i\omega-\epsilon_f-\Sigma_f(\bk,i\omega)-\frac{|V|^2\Delta_\bk^2}{i\omega-\xi_\bk}\right]^{-1}
\en
Next we assume that the self-energy is very weakly dependent on momentum, $\Sigma_f(\bk,i\omega)\simeq\Sigma_f(k_F,i\omega)$
($k_F$ is the conduction electron's Fermi momentum) and expand it to the lowest order in Matsubara frequency:
\beg\label{Sigf}
\Sigma_f(\bk,i\omega)\simeq\Sigma_f(k_F,i\omega)+i\omega\left[\frac{\partial\Sigma_f(k_F,i\omega)}{\partial (i\omega)}\right]_{i\omega\to 0}
\en
Taking into account expressions (\ref{Gff},\ref{Sigf}), for the correlators we find
\beg
\begin{split}
G_{cc}(\bk,i\omega)&=\frac{i\omega-{\varepsilon}_f}{(i\omega-\xi_\bk)(i\omega-{\varepsilon}_f)-|\tilde{V}|^2\Delta_\bk^2}, \\
G_{ff}(\bk,i\omega)&=\frac{i\omega-{\xi_\bk}}{(i\omega-\xi_\bk)(i\omega-{\varepsilon}_f)-|\tilde{V}|^2\Delta_\bk^2},
\end{split}
\en
where $\xi_\bk=-2t\sum_{a=x,y,z}\cos k_a$ is the bare  spectrum of conduction electrons  taken relative to the chemical
potential, $\varepsilon_{f}=Z[\epsilon_{f}+\Sigma_f(0)] $ is the renormalized $f$-level, $\tilde{V}=\sqrt{Z}V$ and $Z=(1-\partial\Sigma_f(k_F,\omega)/\partial\omega)_{\omega=0}^{-1}$. These propagators correspond to the following effective Hamiltonian \cite{Ikeda1996}:
\beg\label{Heff}
\mathcal{H}_{eff}(\bk)=
\left(
\begin{matrix}
\xi_\bk\underline{1} & \tilde{V}{\Phi}_{\Gamma\bk}\dg \\
\tilde{V}{\Phi}_{\Gamma\bk} & \varepsilon_f\underline{1}
\end{matrix}
\right),
\en
Here $\underline{1}$ denotes the unit $2\times2$ matrix.  
The KI is formed if the chemical potential of the quasiparticles lies inside
the hybridization gap, separating the two bands with the spectra
$E_{\pm}(\bk)=
\frac{1}{2}[\xi_\bk+\varepsilon_f\pm\sqrt{(\xi_\bk-\varepsilon_f)^2+4\left|\tilde{V}\Delta_\bk\right|^2}]$.

To discuss the topological properties of our effective model for the KI (\ref{Heff}), 
we need to consider separately the form factors for different $\Gamma$'s. 
It is convenient to distinguish these states according to
their orbital symmetry parameterized by the index $a=1,2,3$ and the
pseudo-spin quantum number ($\alpha=\pm$) \cite{Moreno2000}. Hence, we
have ${f}_{1\pm}\dg|0\rangle=|\pm1/2\rangle$,
${f}_{2\pm}\dg|0\rangle=|\pm3/2\rangle$, and
${f}_{3\pm}\dg|0\rangle=|\pm5/2\rangle$.

The momentum-dependence of the hybridization gap $\Delta_a(\bk)$
follows from Eq. (\ref{eq2}). At small momenta $\bk $, 
$\Delta_1(\bk)=\frac{1}{12}\sqrt{\frac{3}{\pi}}[12\cos(2\theta)+5(3+\cos(4\theta))]^{1/2}$,
$\Delta_2(\bk)=\frac{1}{8}\sqrt{\frac{3}{\pi}}|\sin\theta|[17+15\cos(2\theta)]^{1/2}$,
and $\Delta_3(\bk)=\frac{1}{4}\sqrt{\frac{15}{2\pi}}\sin^2\theta$,
where $\theta$ and $\phi$ define the direction of the unit vector
$\hat{\bf k}$, associated with the point on the Fermi surface. Note
that the hybridization gap has a line of nodes along the $z$-axis for
the shapes $a=2,3$, but generic combinations of all three
form-factors characteristic of contain no nodes. The key results of this Section are 
most simply illustrated using the nodeless $a=1$ Kramers doublet as the ground-state
of the magnetic ion. 
\begin{figure}[h]
\includegraphics[width=3.2in,angle=0]{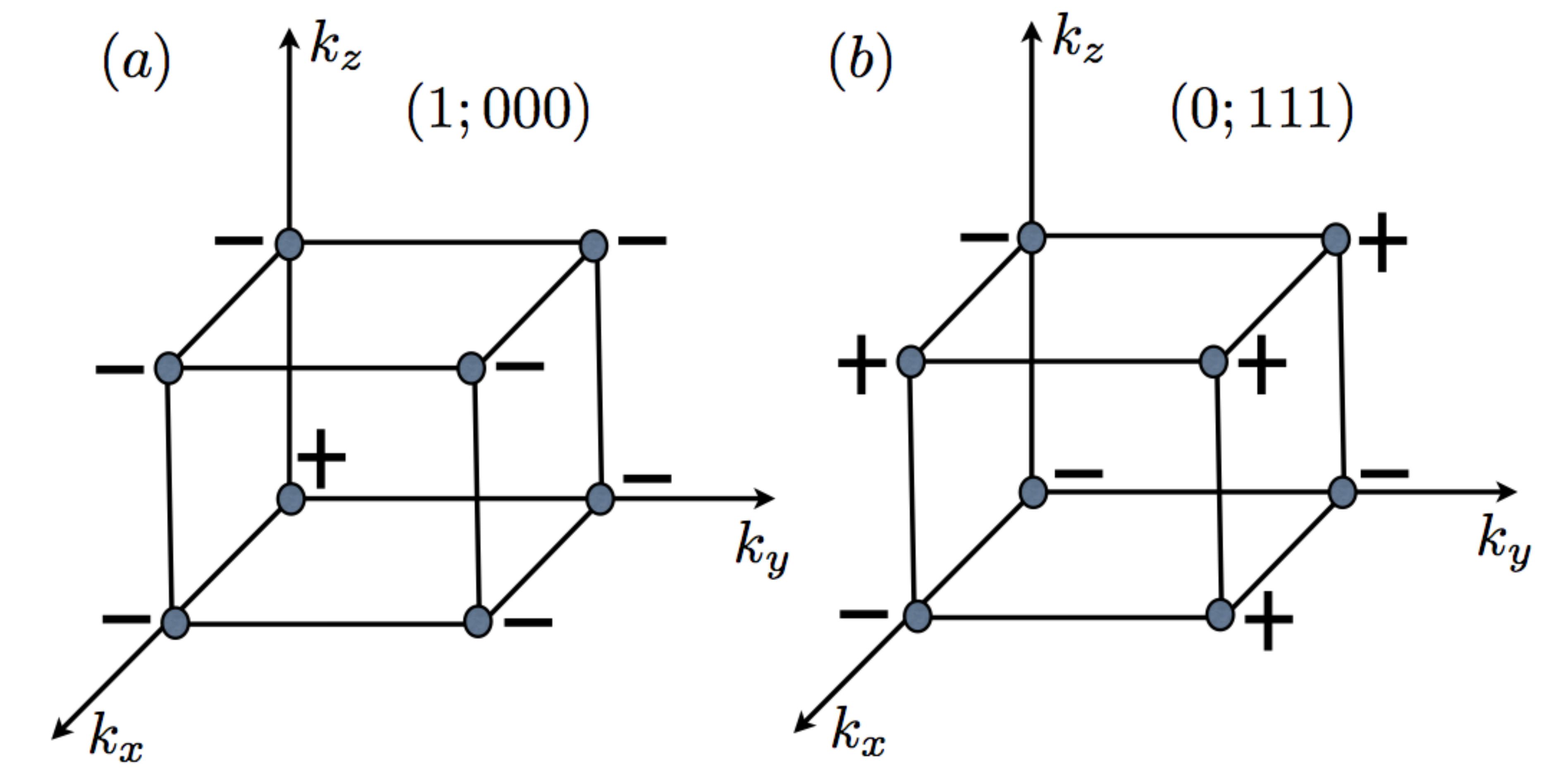}
\caption{Two topological classes can be realized in our model of Kondo insulators for $\varepsilon_f<2t$ (see text). The first 
class with index $\nu=(1;000)$ corresponds to a strong topological insulator and is realized when $\varepsilon_f<-2t$. 
The second class with index $\nu=(0;111)$ is realized for $-2t<\varepsilon<2t$. When the renormalized position of the
$f$-level is at the boundaries, $\varepsilon_f=\pm 2t, \pm 6t$ the system is metallic.}
\label{Fig1}
\end{figure}

To analyze the topology of the bands we  use the fact that 
topology  is invariant under any adiabatic deformation
of the Hamiltonian. We begin our study with a 
tight-binding model for a KI on a simple cubic lattice. Our choice of hybridization ensures that 
the mean-field Hamiltonian (Eq. \ref{Heff}) is
a periodic function satisfying $\mathcal{H}_{eff}({\bf
k})=\mathcal{H}_{eff}({\bf k}+{\bf G})$.
The technical analysis is readily generalized to 
more complicated cases as discussed below.  The most important element 
of the analysis is the odd parity form factor of the 
$f$ electrons, ${\Phi}_{a}(\bk)=-{\Phi}_{a}(-\bk)$.  
This parity property is the only essential input as far as the topological
structure is concerned.

\subsection{calculation of topological indices} 
In Ref.~\onlinecite{FuKane2007}, 
Fu and Kane demonstrate that in an insulator
{\em with time-reversal and space-inversion symmetry}, the topological
structure is determined by parity properties at the eight
high-symmetry points, $\bk^*_m$, in the 3D BZ which are invariant
under time-reversal, up to a reciprocal lattice vector:
$\bk^*_m=-\bk^*_m+{\bf G}$.  In our
case, these symmetries require that $\mathcal{H}_{eff}({\bf k})={P}
\mathcal{H}_{eff}(-{\bf k}){P}^{-1}$ and $\mathcal{H}_{eff}({\bf
k})^{T}={\cal T} \mathcal{H}_{eff}(-{\bf k}){\cal T}^{-1}$, where
the parity matrix $P$
and the unitary part of the time-reversal
operator ${\cal T}$ are given by
\begin{equation}\label{}
P = \begin{pmatrix} \underline{1}& \cr & -\underline{1}\end{pmatrix},
\qquad 
{\cal T} = \begin{pmatrix}  i \sigma_{2}&\cr & i \sigma_{2}  \end{pmatrix},
\end{equation}
where $\sigma_{2}$ is the second Pauli matrix. 
For any space-inversion-odd form factor, it follows immediately that
$\hat{\Phi}_{a}(\bk)=0$ at a high-symmetry point. Hence, the
Hamiltonian at this high symmetry point is simply
$\mathcal{H}_{eff}({\bk^*_m})=(\xi_{\bk^*_m}+\varepsilon_f)
I/2+(\xi_{\bk^*_m}-\varepsilon_f){P}/2$, where $I$ is 
the four-dimensional identity matrix.

The parity at a high symmetry point is thus determined by $\delta_m=\textrm{sgn}(\xi_{\bk^*_m}-\varepsilon_f)$.
Four independent $Z_2$ topological indices $(\nu_0;\nu_1,\nu_2,\nu_3)$ ~\cite{Kitaev}, one strong ($a=0$) and three weak indices ($a={1,2,3}$) can be
constructed from $\delta_m$: (i)~The strong topological index is the product of all eight $\delta_m$'s: 
$I_{\rm STI} = (-1)^{\nu_0}=\prod\limits_{m=1}^{8} \delta_m = \pm 1$; 
(ii)~by setting $k_j=0$ (where $j= x,y, \mbox{and } z$),  
three high-symmetry planes, $P_j = \left\{ {\bf k}: k_j=0\right\}$, are formed that contain four high-symmetry points each. The product of the parities at these four points defines the
corresponding weak-topological index, $I_{\rm WTI}^a =(-1)^{\nu_a}= \prod\limits_{{\bf k}_m \in P_j} \delta_m = \pm 1$, $a=1,2,3$ 
with integers corresponding to the axes $x,y$ and $z$. The existence of the three weak topological indices in 3D is related to a $Z_2$ topological index for 2D systems (a weak 3D TI is similar to a stack of 2D $Z_2$ topological insulators). 
Because there are three independent ways to stack 2D layers to form a 3D system,
the number of independent weak topological indices is also three.
A conventional band insulator has all of the four indices $I_{\rm STI}
= I_{\rm WTI}^x=I_{\rm WTI}^y=I_{\rm WTI}^z = +1$ or  equivalently (0;0,0,0). An index
$I=(-1)$ ($\nu_a=1$) indicates a $Z_2$ topological state with the odd number of
surface Dirac modes. In a KI the symmetry index $\delta_{m}$ of a particular
high symmetry point $m$ is negative provided
$\xi_{\bk^*_{m}}<\epsilon_{f}$ is lower the f-energy $\epsilon_{f}$. 
Thus if 
$\xi_{{\bk_m^*}=0}<\varepsilon_{f}$  at the $\Gamma$ point, while
 $\xi_{\bf{{\bk^*_m\ne 0}}}>\varepsilon_{f}$ for all other symmetry
 points,  then $I_{\rm STI}
= -1$, and hence {the Kondo insulating state is a
strong-topological insulator, robust against disorder} Fig. \ref{Fig1}. Weak-topological insulators and topologically trivial
insulators can in principle be found for different band structures and
different values of $\varepsilon_{f}$. A particularly
interesting possibility is to tune topological phase transitions
between different types of insulators (e.g., by applying a pressure). 
Although we have been specifically considering a tight-binding
model with a primitive unit cell, all our conclusions apply directly
to systems adiabatically connected to this model. 

\begin{figure}[t]
\includegraphics[width=3.0in,angle=0]{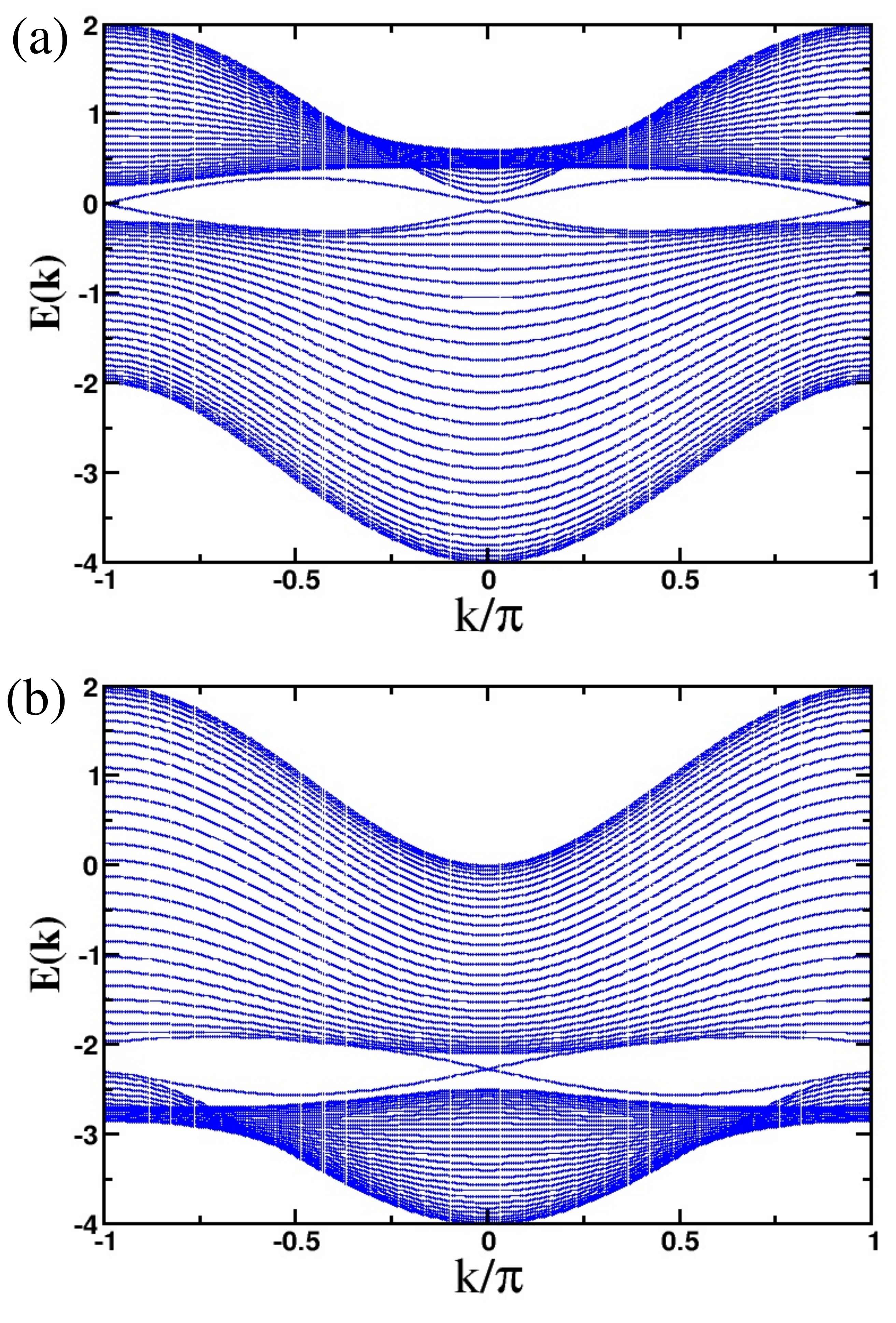}
\caption{Single particle band spectrum governed  by the mean field Hamiltonian (\ref{Heff}) 
along the $x$-axis, $k=k_x, k_y=0$. Top panel shows the 
band structure for the weak topological insulator with the two Dirac points at the Brillouin zone boundaries. 
Band structure for the strong topological insulator with the Dirac point inside the band gap (bottom panel).}
\label{Fig2}
\end{figure}

\subsection{surface state calculation}
In addition to the method discussed
by us in Ref. \cite{Dzero2010}, this can be proven by 
the direct calculation of the band spectrum together with the calculation of the entanglement entropy (see below).
In what follows, we will assume that the $f$-electrons have very weak hole-like dispersion, i.e. 
$\varepsilon_f\to\varepsilon_{f\bk}=2t_f\sum\limits_{i=x,y,z}\cos k_i+\mu_f$, where $t_f=0.1t$ and $\mu_f$ is a chemical 
potential. This implies, in particular, the the boundary separating the WTI and STI are now given by $\mu_{c}'=\pm 2(t+t_f)$
Here, as before, the value of $\mu_c'$ is taken relative to the chemical potential of the conduction electrons.

In order to demonstrate that there is a metallic surface state in the spectrum described by the Hamiltonian 
(\ref{Heff}) we consider a stack of $N=30$ planes along the $z$-direction and diagonalize the Hamiltonian. 
The resulting Hamiltonian matrix has blocks along the diagonal, which describe the hopping and hybridization within
each plane and the off-diagonal parts describing the hopping and hybridization between the planes. I
For the set of the parameters corresponding to the strong topological  Kondo insulator we compute the
spectrum numerically and show the results on Fig. \ref{Fig2}. {For simplicity we have chosen the model form factor, given by:}
\begin{equation}\label{Phi}
\underline{\Phi } = \left\{
\begin{matrix}
V (\sin k_{x}\sigma_{x}+ \sin k_{y}\sigma_{z}), \textrm{ within the planes}, \\
iV_z\sigma_z, ~\textrm{between the planes (upwards)}, \\
-iV_z\sigma_z, ~\textrm{between the planes (downwards)}.
\end{matrix}
\right.
\end{equation}
We see that for the case of strong topological insulator there appears a Dirac point in the gap in the middle
of the Brillouin zone (BZ). For the set of parameters giving a weak topological insulators, there are two Dirac points
located at the edges of the BZ.  We note that the Dirac node in the spectrum exists not only for the simple cubic unit cell, but also for the more complicated fcc- and bcc-unit cells.

\subsection{entanglement entropy and spectrum}
In this Section we independently re-derive our results from the previous subsections by employing the concept of the entanglement 
entropy. In discussions on topological insulators without electron-electron interactions it is implicitly assumed that the presence of the gapless edge modes is a signature of the topologically non-trivial insulating state. In fact, this assumption is confirmed within our description of Kondo insulators. It is interesting, however, to check the topological properties of our model by discussing the properties of the eigenfunctions only. Such an approach has been pioneered by Freedman and collaborators \cite{Freedman} who showed that
topologically nontrivial states of matter can exist without exhibiting the chiral edge modes. In this and the following Section we will discuss in detail the topological properties of the eigenfunctions governed by our effective model Hamiltonian (\ref{Heff}). 

As it has been extensively discussed in the literature (for the more recent accounts see [\onlinecite{Haldane2008,Bernevig2009,Ashvin2009}] and references therein), entanglement entropy can be used to distinguish the topological phases from the non-topological ones. The following criterion is used: the topologically nontrivial state should have non-zero entanglement entropy when the latter can not be tuned to zero by an adiabatic change of the parameters of the system \cite{Bernevig2009}. 

As an aside, we note that our effective Hamiltonian (\ref{Heff}) is a single-particle Hamiltonian and therefore, by calculating its entanglement spectrum we can also test the idea of adiabatic connectivity between our interaction-driven topological Kondo insulators and non-interacting topological insulators. The latter, however, cannot be adiabatically connected to trivial band insulators without making the system gapless. Note, for the trivial insulators we adopt the following definition \cite{Bernevig2009}: upon adiabatic change in the hopping elements to zero, a trivial insulator goes into an atomic insulator without closing the energy gap along the adiabatic path. 

The entanglement entropy can be generally written as 
\beg\label{Sent}
S_{ent}=-\sum\limits_a\left(\xi_a\log\xi_a+(1-\xi_a)\log(1-\xi_a)\right),
\en
where $\{\xi_a\}$ are the single-particle entanglement eigenvalues, subscript $a$ labels the eigenstates.
When the entanglement eigenvalues are neither zero or one, the entanglement entropy is non-zero
and therefore the system is topologically non-trivial. In particular, for translationally invariant topological insulator 
the spacial cut reveals the surface states and yields non zero entanglement entropy \cite{Bernevig2009,Ashvin2009}. 
In this case the entanglement eigenvalues are also labeled by the conserved components of the momentum, 
say, $\xi_a(k_x,k_y)$ for the cut in the $xy$-plane.
Thus the problem of checking whether the insulator is 
topological or not reduces to the problem of determining the entanglement eigenvalues.  
For the case of the Kondo insulators, the computation of the entanglement spectrum may serve as an additional indicator of the nontrivial nature of their ground state especially for the case of complicated lattice structure when the simple approaches for the computation of the $Z_2$ indices do not apply.
The procedure of how these eigenvalues are computed will be given below. 

In this Section we will evaluate the entanglement entropy for the mean field Hamiltonians (\ref{Heff}). 
using the Peschel's method \cite{Bernevig2009,Ashvin2009,Peschel}. The entanglement spectrum 
is determined by correlation function
\beg\label{Gij}
G_{ij}^{\alpha\beta}=\langle\hat{\psi}_{i\alpha}\dg\hat{\psi}_{j\beta}\rangle,
\en
where $\hat{\psi}_{i\alpha}$ creates an electron in state $\alpha=1,...,4$ (conduction or $f$- electron with spin up or down)
on site $i$ and the expectation value is evaluated in the ground state. Introducing the normal operators $\gamma_{n\bk}$, 
where $n$ is the number of the eigenvalues:
\beg
\hat{\psi}_{i\alpha}=\sum\limits_{n=1}^{N_b} e^{i\bk\cdot{\br}_i} u_{n\alpha}(\bk)\hat{\gamma}_{n\bk},
\en
where $N_b=2$ is the number of the occupied bands and $u_{n\alpha}(\bk)$ are the eigenvectors. For the correlation function we find
\beg
G_{ij}^{\alpha\beta}=\sum\limits_{\bk}e^{i\bk\cdot(\br_i-\br_j)}\sum\limits_{n=1}^{2}u_{n\alpha}^*(\bk)u_{n\beta}(\bk)
\en
summation goes over all components of the momentum $\bk$. 

Let us imagine now that our system is cut in two halves along a given 
spacial directions. To be specific, let us make the cut along the $xy$-plane, so that $k_x$ and $k_y$ are conserved. 
The entanglement spectrum $\xi_{a}(\bk_\perp)$, $\bk_\perp=(k_x,k_y)$ Eq. (\ref{Sent}), will then be given by the eigenvalues of the following matrix:
\beg
G_{ij}^{\alpha\beta}(k_x,k_y)=\sum\limits_{k_z}e^{ik_z\cdot(z_i-z_j)}\sum\limits_{n=1}^{2}u_{n\alpha}^*(\bk)u_{n\beta}(\bk)
\en
where $i,j$ are confined to the right (or left) part of the system. Specifically, we need to solve the following eigenvalue problem:
\beg
\sum\limits_{j,\beta}G_{ij}^{\alpha\beta}(\bk_\perp)\varphi_{j\beta}^{(a)}(\bk_\perp)=\xi_a(\bk_\perp)\varphi_{i\alpha}^{(a)}(\bk_\perp)
\en

\begin{figure}[h]
\includegraphics[width=2.7in,angle=0]{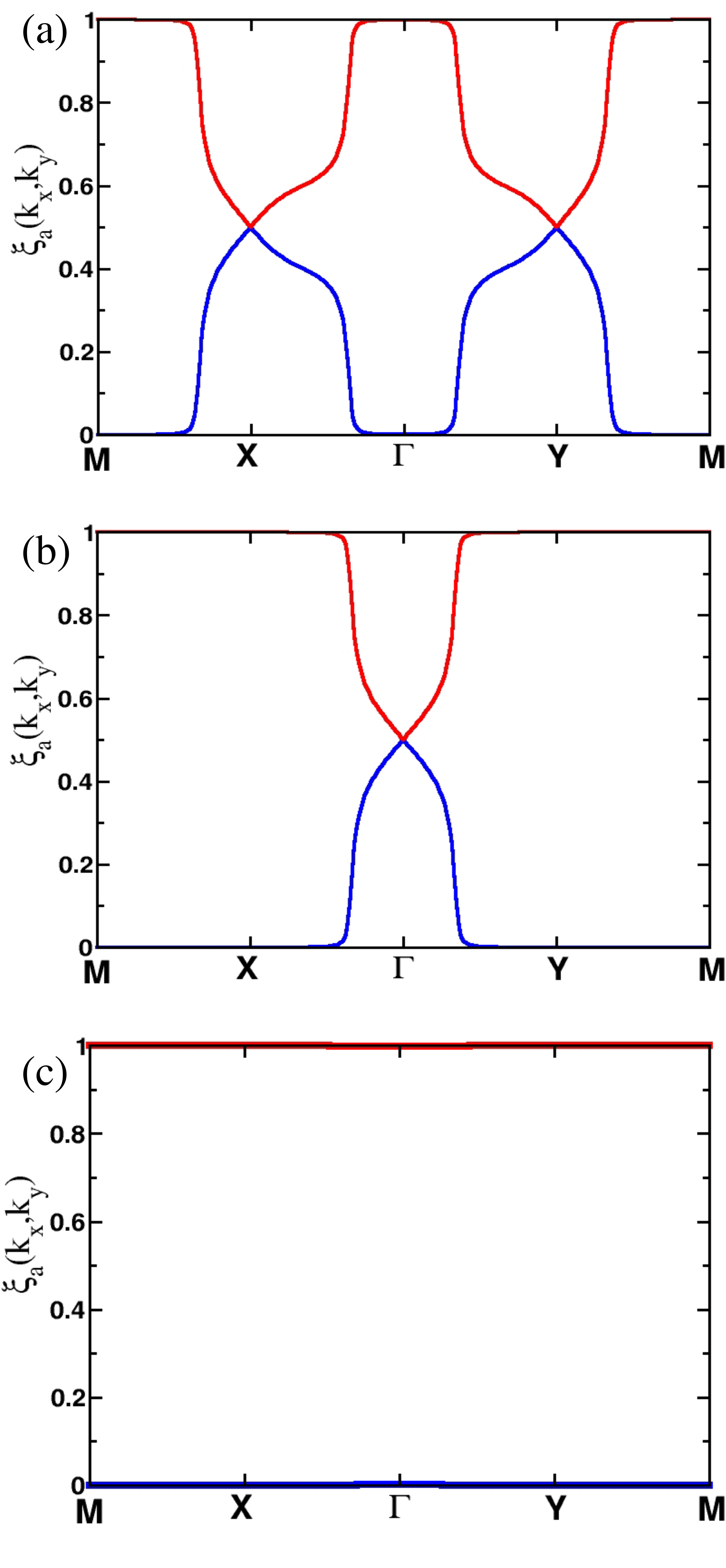}
\caption{Entanglement spectrum for the model of Ce-based Kondo insulator: (a) weak topological insulator $(0;111)$, (b) strong topological insulator $(1;000)$ and (c) trivial insulator $(0;000)$. For the presentation purposes we choose only two eigenvalues for each momentum $\bk_\perp$. }
\label{Fig3}
\end{figure}

\begin{figure*}
\includegraphics[width=6.7in,angle=0]{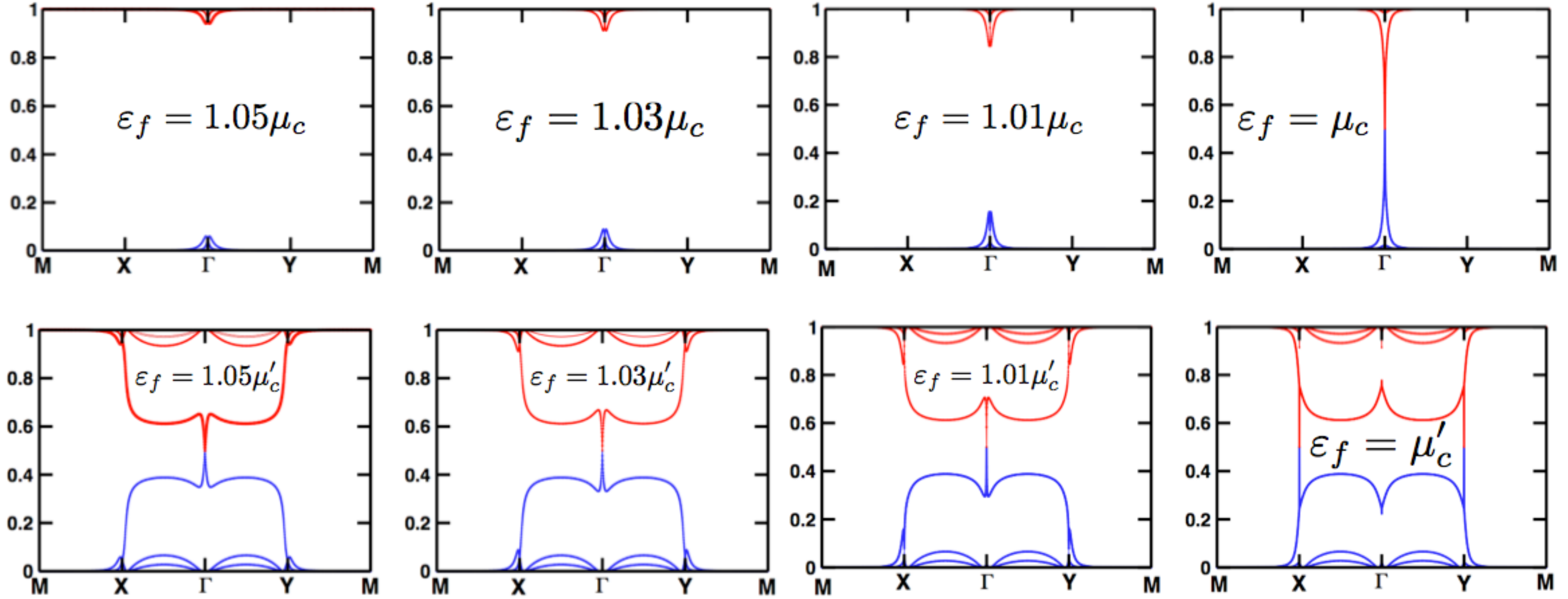}
\caption{Evolution of the entanglement spectrum with the change in the position of the renormalized $f$-level, $\varepsilon_f$.
Top four panels show the change in the entanglement spectrum as the system goes from the trivial insulator to the strong topological insulator. Bottom four panels illustrate the changes in spectrum as system goes from strong to weak topological insulator. Note that 
when $\varepsilon_f$ is exactly at the boundary separating different insulating phases so that the bulk insulating gap vanishes, the vertical lines in the entanglement entropy spectrum reflect the absence of the surface modes at the boundaries.}
\label{FigEvo}
\end{figure*}

We show the results of our computation for the model Hamiltonian (\ref{Heff}) with $N_b=2$ on Figs. \ref{Fig3} and \ref{FigEvo}. As we can see, depending on the position of renormalized $f$-level relative to the bottom of the conduction band, we find either singe node or two nodes in the entanglement spectrum. The number of the nodes is equal to the number of nodes of the surface states inside 
the insulating gap, in complete agreement with our expectations. 

To summarize,  our results from this Section confirm that $Z_2$-odd topological Kondo insulators cannot be adiabatically connected to $Z_2$-even
insulators adiabatically without vanishing of the insulating gap along the adiabatic path. At the same time we see one-to-one correspondence between the Kondo insulators and non-interacting $Z_2$ topological insulators confirming the idea of adiabatic connectivity between  the two discussed in the Introduction.

\section{Construction of the Wannier wave functions}
Within our model of Kondo insulators we can also address the problem of constructing maximally localized Wannier 
functions (WF). This question has several important applications in the general theory of topological insulators \cite{Vander2010,Yu2011}, such as calculation of the $Z_2$ indices as well as characterization of the topological structure using first principles calculations. 

Generally, the construction of the WF proceeds in two stages. The first stage has to do with the initial choice of the basis set before
specifying a particular choice of the gauge. 
This needs be done in order to make the WF nonsingular across the whole Brillouin zone. The gauge is then fixed by imposing the certain criterion. As an example, maximum localization criterion is typically used \cite{Marzari1997}.
Apart from the problem related to the arbitrariness in the choice of Wannier functions, there exists a topological obstruction for constructing the Wannier functions for Chern insulators realized in systems with broken time-reversal symmetry \cite{Timo,Thouless}. 
As it turns out, in the case of the $Z_2$ topological insulators there is also a topological obstruction albeit a less severe one. As it was recently discussed in Ref. [\onlinecite{Vander2010}] for the Kane-Mele model, it is impossible to construct the time-reversal invariant basis set of the Wannier functions, but one can construct the basis set consisting of the non-Kramers pairs. The above mentioned arbitrariness in the definition of the WF is then fixed by the criterion of maximum localization \cite{Yu2011,Marzari1997}. 

In this section we will specifically apply the prescription developed in Ref. [\onlinecite{Vander2010}] to construct the basis
set which then can be used to initialize the procedure to compute the maximum localized WF. On one hand, this should provide another example of the manifestation of above mentioned obstruction and the way it can be resolved. On the other hand, it gives 
an insight into the structure of the wave functions describing the quasiparticles in the occupied bands. 

\subsection{preliminaries}
Below we will follow almost verbatim the discussion in Refs. [\onlinecite{Vander2010,Marzari1997}].
For variety of applications (i.e. numerical calculations) it is required that the Bloch-like wave functions must remain smooth across the whole Brillouin zone (BZ). The problem is that the specific choice of the Bloch functions 
$|\psi_{n\bk}\rangle$ is not unique, since these wave functions  
have an additional gauge freedom originating from possibility of mixing 
with the wave functions describing the occupied bands:
\beg\label{gauge}
|\psi_{n\bk}\rangle\to\sum\limits_{m}{\cal U}_{nm}(\bk)|\psi_{m\bk}\rangle
\en
(here the summation goes over the occupied bands). For all the practical purposes, however, the freedom of choosing the proper gauge transformation must be removed by applying some restrictions on choosing the specific gauge.
The latter uses the criterion of maximum localization of the WF \cite{Marzari1997}. WFs are defined by
\beg\label{WFs}
W_n(\bfr-{\bf R})=\frac{\Omega}{(2\pi)^3}\int_{BZ}e^{-i\bk\cdot{\bf R}}\psi_{n\bk}(\bfr),
\en
where $\Omega$ is a volume of the unit cell and $\psi_{n\bk}(\bfr)=\langle\bfr|\psi_{n\bk}\rangle$ are the Bloch 
wave functions, $n$ is a band index and ${\bf R}$ is a position of a lattice site. 

The unitary transformation (\ref{gauge}) can be initialized using the following procedure. 
One first chooses the set of localized \emph{trial} wave functions
$|\tau_{i\bk}\rangle$ and then form a set of new basis functions
\beg\label{eta}
|\tilde{\tau}_{i\bk}\rangle=\sum\limits_{n=1}^{\cal N}|\psi_{n\bk}\rangle\langle\psi_{n\bk}|\tau_{i\bk}\rangle, \quad i=1,{\cal N}
\en
where ${\cal N}$ is the number of the occupied bands. Since this new basis set is not orthonormal, one can adopt a L\"{o}wdin procedure and form the overlap matrix
\beg\label{Smn}
S_{mn}(\bk)=\langle\tilde{\tau}_{m\bk}|\tilde{\tau}_{n\bk}\rangle.
\en
Now we can use Eqs. (\ref{eta},\ref{Smn}) to form a set of Bloch-like states
\beg\label{tpsi}
|\widetilde{\psi}_{n\bk}\rangle=\sum\limits_{m}\left[S^{-1/2}(\bk)\right]_{mn}|\tilde{\tau}_{m\bk}\rangle
\en
These states, albeit not eigenstates of the Hamiltonian, should be the smooth functions of the quasi-momentum
$\bk$ and are used to construct the localized set of the WFs:
\beg
\widetilde{W}_n(\bfr-{\bf R})=\frac{\Omega}{(2\pi)^3}\int_{BZ}e^{-i\bk\cdot{\bf R}}\widetilde{\psi}_{n\bk}(\bfr)
\en
The above construction breaks of the determinant of the matrix $S_{mn}(\bk)$ vanishes in some points
of the BZ. Thus the problems consists in finding the proper set of trial states (\ref{eta}) such that
det$[S(\bk)]\not=0$. Finally, we note that required degree of localization can be achieved by employing the
iterative procedure \cite{Marzari1997}.

We now construct the Wannier functions for our mean field model described by the Hamiltonian (\ref{Heff}).
For the Bloch wave functions we write
\beg\label{psink}
|\psi_{n\bk}\rangle=\sum\limits_{s=1}^4C_{sn\bk}|{s\bk}\rangle,
\en
where $n=1,2$ labels the occupied bands, coefficients $C_{sn\bk}$ are the components of the eigenvectors of the 
Hamiltonian (\ref{Heff}) 
and the summation goes over the components of generalized spinor which includes spinfull conduction (c) and f-elecron (f) states: 
\beg\label{skdef}
\begin{split}
|{s=1,\bk}\rangle=\hat{c}_{\bk\up}\dg|0\rangle, ~|{s=2,\bk}\rangle=\hat{c}_{\bk\dn}\dg|0\rangle, \\
|{s=3,\bk}\rangle=\hat{f}_{\bk\up}\dg|0\rangle, ~|{s=4,\bk}\rangle=\hat{f}_{\bk\dn}\dg|0\rangle
\end{split}
\en
In Eq. (\ref{psink}) the basis functions $|{s\bk}\rangle$ are defined on the each site on the lattice $\bR$, i.e. 
\beg\label{skr}
|s\bk\rangle=\frac{|s\rangle}{\sqrt{N}}\sum\limits_{\br}e^{i\bk\cdot\br}\delta(\br-\bR)
\en
In what follows we adopt the method outlined above to construct the Wannier functions for our model Kondo insulators. 
\subsection{choice of the basis}
\begin{figure}[h]
\includegraphics[width=3.0in,angle=0]{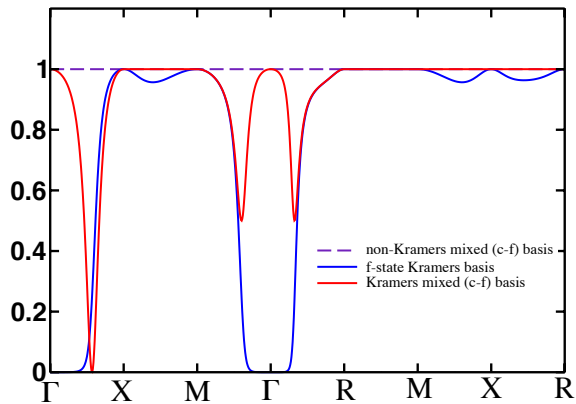}
\caption{Plot of the dependence of $\textrm{det}[\hat{S}(\bk)]$ along the path in the BZ.
The elements of the matrix $\hat{S}(\bk)$ has been obtained using the trial basis set, which consists of 
(a) non-Kramers pair of states each containing the superposition between the conduction and $f$-states; (b) Kramers pair of $f$-states and (c) Kramers pair states with linear superposition of conduction and $f$-electron wave functions. Determinant does not
vanish anywhere in the BZ only for the basis (c).}
\label{Fig4}
\end{figure}

Onset of the coherence in the Kondo lattice can be interpreted as an emergence of new quasi-particles which are the 
linear superposition of the localized and conduction states. Since the newly formed quasiparticle band is narrow, the 
spectral weight is mostly governed by the $f$-states. Thus, to construct the Wannier functions we first consider the basis
on $f$-states only:
\beg\label{choice1}
\begin{split}
|\tau_{1\bk}\rangle=|3\bk\rangle, \quad |\tau_{2\bk}\rangle={|4\bk\rangle}
\end{split}
\en
(see Eqs. (\ref{skdef},\ref{skr})).
For the new set of basis vectors (\ref{eta}) with the help of Eqs. (\ref{psink},\ref{choice1}) this implies
\beg
\begin{split}
|\tilde{\tau}_{1\bk}\rangle&=C_{31\bk}^*|\psi_{1\bk}\rangle+C_{32\bk}^*|\psi_{2\bk}\rangle, \\
|\tilde{\tau}_{2\bk}\rangle&=C_{41\bk}^*|\psi_{1\bk}\rangle+C_{42\bk}^*|\psi_{2\bk}\rangle, \\
\end{split}
\en
For the determinant of the matrix $\hat{S}(\bk)$ we find
\beg
\begin{split}
\textrm{det}[\hat{S}(\bk)]=&(|C_{31\bk}|^2+|C_{32\bk}|^2)(|C_{41\bk}|^2+|C_{42\bk}|^2)\\&-
|C_{31\bk}C_{41\bk}^*+C_{32\bk}C_{42\bk}^*|^2
\end{split}
\en
We present the results on Fig. \ref{Fig4}. We see that the determinant of the matrix (\ref{Smn}) is zero near the $\Gamma$-point which means
that the choice (\ref{choice1}) is not suitable for construction of non-singular Bloch functions and consequently Wannier functions. 
The same result holds for the trial basis built out of the conduction states, $|1\bk\rangle$ and $|2\bk\rangle$ as well as their linear combinations.  

As we have mentioned above, formation of the coherence in the Kondo lattice can be seen as a formation of the new states
(or quasiparticles) as a result of the hybridization between the conduction and $f$ electrons. 
Motivated by this observation, let us try the following two trial basis wave-functions:
\beg\label{tauk}
\begin{split}
|\tau_{1\bk}\rangle&=\frac{1}{\sqrt{2}}(|1\bk\rangle+|3\bk\rangle), \\
|\tau_{2\bk}\rangle&=\frac{1}{\sqrt{2}}(|2\bk\rangle-|4\bk\rangle), \\
\end{split}
\en
Note that the trial basis functions \underline{do not} transform into each other by time reversal operator, so they do not 
form a Kramers doublet. If follows
\beg\label{nonKramers}
\begin{split}
|\tilde{\tau}_{1\bk}\rangle&=\frac{(C_{11\bk}^*+C_{31\bk}^*)}{\sqrt{2}}|\psi_{1\bk}\rangle+
\frac{(C_{12\bk}^*+C_{32\bk}^*)}{\sqrt{2}}|\psi_{2\bk}\rangle, \\
|\tilde{\tau}_{2\bk}\rangle&=\frac{(C_{21\bk}^*-C_{41\bk}^*)}{\sqrt{2}}|\psi_{1\bk}\rangle+
\frac{(C_{22\bk}^*-C_{42\bk}^*)}{\sqrt{2}}|\psi_{2\bk}\rangle, \\
\end{split}
\en
The determinant of the matrix $\hat{S}(\bk)$ up to the numerical pre-factor is
\beg
\begin{split}
\textrm{det}[\hat{S}(\bk)]=&
(|C_{11\bk}+C_{31\bk}|^2+|C_{12\bk}+C_{32\bk}|^2)\times\\&(|C_{21\bk}-C_{41\bk}|^2+|C_{22\bk}-C_{42\bk}|^2)-\\&
-\vert(C_{11\bk}^*+C_{31\bk}^*)(C_{21\bk}-C_{41\bk})+\\&+(C_{12\bk}^*+C_{32\bk}^*)(C_{22\bk}-C_{42\bk})\vert^2
\end{split}
\en
We present the resulting dependence $\textrm{det}[\hat{S}(\bk)]$ on momentum on Fig.~\ref{Fig4}. As we have expected, 
the determinant does not vanish anywhere within the BZ which means we have succeeded in constructing the wave
functions $\vert\tilde{\psi}_{n\bk}\rangle$ (\ref{tpsi}). In fact, we find $\textrm{det}[\hat{S}(\bk)]=1$ for the non-Kramers basis set 
(\ref{nonKramers}). Finally, we note that in agreement to the results of Ref. [\onlinecite{Vander2010}] obtained for the $Z_2$-odd phase in the Kane-Mele model, here our non-singular basis set also consists of the non-Kramers pair of states. To summarize, we have demonstrated that the basis for the Bloch wave functions can be chosen in such a way that no singularities are generated 
across the Brillouin zone. 

\section{conclusions}
In this paper we have discussed the conditions for the emergence of chiral surface states in semiconducting
$f$-electron systems. 
We considered an insulating state in heavy fermion systems which appears at finite temperatures as a result of strong interaction between the conduction and the predominantly localized $f$-electrons. 
Having started with the periodic Anderson lattice model, we considered the low-energy version of that model, which 
takes into account the effect of Hubbard repulsion between the $f$-electrons on the level of renormalizations to the 
$f$-electron energy and hybridization amplitudes. The key ingredient 
of our model is momentum dependent hybridization amplitudes. 
The momentum dependence of the amplitudes originates from the strong spin-orbit coupling interaction on $f$-sites. 
The analysis of the topological structure of the newly formed insulating state is greatly simplified for the systems with 
simple cubic unit cell. In that case, the form factors vanish at high symmetry points of the BZ. This embeds the topological 
singularities into the valence band, so that when the form-factors have $p$- or $f$-wave symmetry it immediately leads
to the topological insulator. 

To describe the physics of Ce-based Kondo insulators, we considered the simplest model containing single conduction band hybridized with the Kramers doublet of $f$-states. We find that there will always be chiral surface states, when hybridization gap does not have nodes. The robustness of these states with respect to disorder is determined by the position of the renormalized $f$-level relative to the bottom of the conduction band. We then verify our results for both models by calculating the entanglement entropy spectrum. Finally, we also discuss how to choose the basis for constructing Wannier wave functions, which are well defined everywhere in the Brillouin zone. It is interesting to note that the required basis relies on superposition between the conduction and the localized $f$-states. More importantly, this agrees with common view that a heavy quasiparticle is the quantum many-body superposition of conduction and $f$-states. 

\section{acknowledgments}
We would like to thank D. Vanderbilt, J. Allen and M. Aronson for stimulating discussions. 
This work was supported by the Ohio Board of Regents Research Incentive Program grant OBR-RIP-220573 (M.D.), JQI-NSF-PFC (K. S.), DOE grant DE-FG02-99ER45790 (P. C.), and NSF-CAREER (V.G.). This work was supported in part by the National 
Science Foundation under grant No. 1066293 and the hospitality of the Aspen Center for Physics.


\begin{thebibliography}{99}

\bibitem{Hasan2010}M. Z. Hasan and C.L. Kane, Rev. Mod. Phys. {\bf 82}, 3045 (2010).

\bibitem{Qi2010}X.-L. Qi and S.-C. Zhang, Rev. Mod. Phys.  {\bf 83}, 1057 (2011).

\bibitem{Fu2007}L. Fu, C. L. Kane and E. J. Mele, Phys. Rev. Lett. {\bf 98}, 106803 (2007).

\bibitem{Moore2007}J. E. Moore and L. Balents, Phys. Rev. B {\bf 75}, 121306(R) (2007).

\bibitem{Roy2009} R. Roy, Phys. Rev. B {\bf 79}, 195322 (2009).

\bibitem{Hsieh2008} D. Hsieh, D. Qian, L. Wray, Y. Xia, Y. S. Hor, R. J. Cava and M. Z. Hasan, Nature {\bf 452}, 970 (2008).

\bibitem{Xia2009}Y. Xia, D. Qian, D. Hsieh, L.Wray, A. Pal, H. Lin, A. Bansil,
D. Grauer, Y. S. Hor, R. J. Cava and M. Z. Hasan, Nat. Phys. {\bf 5}, 398 (2009).

\bibitem{exp1} P. Roushan, J. Seo, C. V. Parker, Y. S. Hor, D. Hsieh, D. Qian, A. Richardella, M. Z. Hasan, R. J. Cava and A. Yazdani, Nature {\bf 460}, 1106 (2009).

\bibitem{exp2} T. Zhang, P. Cheng, X. Chen, J. F. Jia, X. C. Ma, K. He, L. L. Wang, H. J. Zhang, X. Dai, Z. Fang, X. C. Xie and Q. K. Xue, Phys. Rev. Lett. {\bf 103}, 266803 (2009).

\bibitem{exp3} J. Seo, P. Roushan, H. Beidenkopf, Y. S. Hor, R. J. Cava, and A. Yazdani, Nature {\bf 466}, 343 (2010).

\bibitem{exp4} Z. Alpichshev, J. G. Analytis, J. H. Chu, I. R. Fisher, Y. L. Chen, Z. X. Shen, A. Fang and A. Kapitulnik, 
Phys. Rev. Lett. {\bf 104}, 016401 (2010).

\bibitem{Raghu2008} S. Raghu, X.-L. Qi, C. Honerkamp, S.-C. Zhang, Phys. Rev. Lett. {\bf 100}, 156401 (2008).

\bibitem{Sun2009} K. Sun, H. Yao, E. Fradkin and S. A. Kivelson, Phys. Rev. Lett. {\bf 103}, 046811 (2009).

\bibitem{Nandkishore2010} R. Nandkishore and L. Levitov Phys. Rev. B {\bf 82}, 115124 (2010).

\bibitem{Sun2010} K. Sun, W. Vincent Liu and S. Das Sarma, e-print arXiv:1011.4301.

\bibitem{pyro1} H. M. Guo and M. Franz, Phys. Rev. Lett. {\bf 103},  206805 (2009).

\bibitem{pyro2} D. A. Pesin and L. Balents, Nat. Phys. {\bf 6}, 376 (2010).

\bibitem{pyro3} X. A. Wan, A. Turner, A. Vishwanath and S. Y. Savrasov,  e-print arXiv: 1007.0016.

\bibitem{pyro4} B.-J. Yang and Y. B. Kim,  e-print arXiv:1004.4630.

\bibitem{Dzero2010} M. Dzero, K. Sun, V. Galitski and P. Coleman, Phys. Rev. Lett. {\bf 104}, 106408 (2010).


\bibitem{skuter} Binghai Yan, Lukas M\"{u}chler, Xiao-Liang Qi, Shou-Cheng Zhang and Claudia Felser, 
 e-print arXiv:1104:0641.

\bibitem{Chazalviel1976} J. N. Chazalviel, M. Campagna, G. K. Wertheim, and P. H. Schmidt, 
Phys. Rev. B {\bf 14}, 4586 (1976). 

\bibitem{Geballe69}A. Menth, E. Buehler and T. H. Geballe, 
Phys. Rev. Lett. {\bf 22}, 295 (1969).

\bibitem{KIReviews1} G. Aeppli and Z. Fisk, Comm. Condens. Matter Phys. {\bf 16}, 155 (1992).

\bibitem{KIReviews2} H. Tsunetsugu, M. Sigrist and K. Ueda, Rev. Mod. Phys. {\bf 69}, 809 (1997).

\bibitem{KIReviews3} P. Riseborough,  Adv. Phys. {\bf 49}, 257 (2000).

\bibitem{Coleman2007} P. Coleman, \emph{``Heavy Fermions: Electrons at the Edge of Magnetism"}, 
Handbook of Magnetism and Advanced Magnetic Materials, 
Vol 1, 95-148 (Wiley, 2007).

\bibitem{allen} R. Martin and J. Allen, J. Appl. Phys {\bf 50}, 7561 (1979).

\bibitem{wang2011} X. Wan, A. M. Turner, A. Vishwanath and S. Y. Savrasov, Phys. Rev. B {\bf 83}, 205101 (2011).

\bibitem{Kikoin1994} K. A. Kikoin, A. de Visser, K. Bakker, T. Takabatake, Z. Phys. B {\bf 94}, 79 (1994).

\bibitem{Ikeda1996} H. Ikeda and K. Miyake, Jour. of Phys. Soc. of Japan {\bf 65}, 1769 (1996).

\bibitem{Moreno2000} Juana Moreno and P. Coleman, Phys. Rev. Lett. {\bf 84}, 342 (2000).

\bibitem{Rob} E. D. Bauer,  A. D. Christianson, J. M. Lawrence, E. A. Goremychkin, N. O. Moreno, N. J. Curro, F. R. Trouw, J. L. Sarrao,    J. D. Thompson, R. J. McQueeney, W. Bao and R. Osborn, J. Appl. Phys. {bf 95}, 7201 (2004).

\bibitem{Coqblin1969} B. Coqblin and J. R. Schriefer, Phys. Rev. {\bf 185}, 847 (1969).

\bibitem{Flint2008} Rebecca Flint, M. Dzero, and P. Coleman, Nature Physics {\bf 4}, 643 (2008).

\bibitem{Hundley1990} M. F. Hundley, P. C. Canfield, J. D. Thompson, and Z. Fisk Phys. Rev. B {\bf 42}, 6842 (1990).

\bibitem{CeNiSn2002} T. Terashima, C. Terakura, S. Uji, H. Aoki, Y. Echizen and T. Takabatake, 
Phys. Rev. B {\bf 66}, 075127 (2002).


\bibitem{FuKane2007} L. Fu and C. L. Kane, Phys. Rev. B {\bf 76}, 045302 (2007).

\bibitem{Kitaev} A. Kitaev, e-print arXiv:0901.2686v2 (2009).


\bibitem{Freedman} M. Freedman, C. Nayak, K. Shtengel and K. Walker, Ann. Phys. {\bf 310}, 428 (2004). 

\bibitem{Haldane2008} H. Li and F. D. M. Haldane, Phys. Rev. Lett. {\bf 101}, 010504 (2008).

\bibitem{Ashvin2009} Ari M. Turner, Yi Zhang and Ashvin Vishwanath, e-print arXiv:0909:3119.

\bibitem{Bernevig2009} T. Hughes, E. Prodan and B. A. Bernevig, e-print arXiv:1010.4508.

\bibitem{Peschel} I. Peschel, J. Stat. Mech. {\bf 10}, P06004 (2004).


\bibitem{Vander2010} A. Soluyanov and D. Vanderbilt, e-print arXiv:1009:1415.

\bibitem{Yu2011} R. Yu, X. L. Qi, A. Bernevig, Z. Feng and X. Dai, preprint arXiv:1101.2011 (2011).

\bibitem{Marzari1997} N. Marzari and D. Vanderbilt, Phys. Rev. B {\bf 56}, 12847 (1997).

\bibitem{Timo} T. Thonhauser and D. Vanderbilt, Phys. Rev. B {\bf 74}, 235111 (2006).

\bibitem{Thouless} J. D. Thouless, J. Phys. C {\bf 17}, 235 (1984). 

\end{thebibliography}
\end{document}